\documentclass[a4paper,11pt]{article}
\pdfoutput=1 

\usepackage{jcappub} 

\usepackage[T1]{fontenc} 

\usepackage{amsmath, amsfonts, amsthm, amssymb, graphicx,comment}

\usepackage{hyperref,subfigure}

\allowdisplaybreaks[3]

\newcommand{\be} {\begin{equation}}
\newcommand{\ee} {\end{equation}}
\def \bal#1\eal  {\begin{align} #1 \end{align}}
\newcommand{\nn} {\nonumber\\}

\newcommand{\ud} {\mathrm{d}}

\newcommand{\nd} {\nabla}
\newcommand{\pd} {\partial}

\newcommand{\bfk} {{\bf k}}

\newcommand{\mc} {\mathcal}

\title{\huge Gravitational waves from Affleck-Dine condensate fragmentation}

\author[a,b]{Shuang-Yong Zhou}
\emailAdd{sxz353@case.edu}
\affiliation[a]{Department of Physics, Case Western Reserve University, 10900 Euclid Ave, Cleveland, OH 44106, USA}
\affiliation[b]{SISSA, Via Bonomea 265, 34136, Trieste, Italy and INFN Sezione di Trieste, Italy}

\date{\today}

\abstract{
We compute the stochastic gravitational wave production from Affleck-Dine condensate fragmentation in the early universe, focusing on an effective potential with a logarithmic mass correction that typically arises in gravity mediated supersymmetry breaking scenarios. We find that a significant gravitational wave background can be generated when Q-balls are being formed out of the condensate fragmentation. This gravitational wave background has a distinct multi-peak power spectrum where the trough is closely linked to the supersymmetry breaking scale and whose frequencies are peaked around kHz for TeV supersymmetry breaking. 
}

\begin{document}

\maketitle
\flushbottom

\section{Introduction}

In the early universe, some scalar fields may develop large vevs, forming Affleck-Dine (AD) condensates  \cite{Affleck:1984fy}. This is generically expected in supersymmetric (SUSY) extensions of the Standard Model, as there are typically plenty of flat directions where the scalar potential vanishes. The Minimal Supersymmetric Standard Model already has hundreds of flat directions \cite{Dine:1995kz, Gherghetta:1995dv}, which leads to interesting cosmological consequences  (see \cite{Dine:2003ax, Allahverdi:2012ju, Enqvist:2003gh} for recent reviews).

For low energy SUSY, the flat directions are lifted by soft breaking terms not far away from the weak scale. These soft terms are negligible in the very early universe when the Hubble parameter $H$ is much larger than the weak scale. Indeed, the finite energy density of the universe can induce an  $\mc{O}(H)$ effective mass for flat directions \cite{Dine:1995uk, Dine:1995kz}. The induced mass term can sometimes be negative, which efficiently drives the field away from the zero value during inflation, and, once non-renormalizable terms are included, the field can be stabilized at a finite value, forming the AD condensate \cite{Dine:1995uk, Dine:1995kz}.  The AD condensate almost adiabatically follows the decrease of $H$, and, when $H$ is surpassed by the soft mass scale, the condensate starts oscillating around the potential's minimum and fragments. Therefore, the scale of AD condensate fragmentation is closely linked to the scale of SUSY breaking. On the other hand, if an AD field also carries the baryon ($B$)/lepton ($L$) number,  the observed matter-antimatter asymmetry can be created in this process \cite{Affleck:1984fy}.

Q-balls are a type of non-topological soliton, which has a localized, time-dependent and non-dissipative field configuration \cite{Coleman:1985ki, fls76}. They exist in a scalar field theory endowed with a continuous global symmetry and an ``open-up'' potential \cite{Coleman:1985ki}. Due to the global symmetry, a Noether charge $Q$ can be defined, and thus a Q-ball is a concentrated lump of $Q$ charges. Physically, they can form because the spherically symmetric Q-ball configuration is energetically favored in such a field theory \cite{Coleman:1985ki}. (See \cite{Copeland:2014qra} for the existence of long-lived composite, charge-swapping Q-balls in the same theory where Q-balls exist.) Q-balls are relevant in the early universe and can be an important consequence of SUSY flat directions (\cite{Kusenko:1997vp, Kusenko:1997zq, Dvali:1997qv, Kusenko:1997si, Enqvist:1997si, Enqvist:1998en, Enqvist:2000gq, Pawl:2004vi, Kasuya:2000sc, Kasuya:1999wu, Kusenko:2008zm, Enqvist:2000cq, Multamaki:2002hv, Copeland:2009as, Campanelli:2007um, Enqvist:2002rj, Kasuya:2007cy, Kusenko:2009cv, Chiba:2009zu,Chiba:2010ff,Kasuya:2008xp} and see \cite{Dine:2003ax, Enqvist:2003gh, Allahverdi:2012ju} for recent reviews).
Indeed, resonance often takes place in the decay process of an AD condensate: Fluctuations in the AD condensate are usually rapidly amplified by the condensate's coherent oscillation, which fragments the AD condensate into Q-balls.

Unlike photos, gravitons decouple from the hot, dense Big Bang plasma at a very high temperature, thus relic gravitational waves (GWs) can be a probe into the very early universe. Significant stochastic GW backgrounds can be generated in early universe phase transitions where resonance takes place, such as in preheating after inflation (see \cite{Khlebnikov:1997di, Easther:2006gt, Easther:2006vd, Giblin:2014dea, GarciaBellido:2007dg,GarciaBellido:2007af, Dufaux:2007pt, Price:2008hq, Dufaux:2008dn, Ashoorioon:2013oha, Ashoorioon:2015hya, Zhou:2013tsa} and references therein).  This also happens in AD condensate fragmentation when the condensate has become dominant in the universe's energy budget before it decays \cite{Kusenko:2008zm, Kusenko:2009cv}. In some cases, spherically symmetric and energetic lumps, such as oscillons and Q-balls, can emerge as the resonance product in the phase transition. As spherical symmetric configurations do not emit GWs, we can expect that, in these cases, the GW frequency bands associated with the lump's characteristic sizes are relatively suppressed, so the GW power spectrum generated has a distinct multi-peak structure. For example, this has been shown to be the case in oscillon preheating after axion monodromy inflation \cite{Zhou:2013tsa}.

In this paper, we calculate the stochastic GW background generated when an AD condensate fragments into Q-balls. Our numerical results will be presented for the running mass model (see Eq.~(\ref{poten})), which typically arises in gravity mediated supersymmetry breaking scenarios. The GW production of this model has been investigated with a lattice computation previously in Ref \cite{Kusenko:2009cv}, but we re-visit it using an independent code and with much higher numerical resolution. We show that the GW power spectrum generated indeed has a multi-peak structure, with the frequency bands of the troughs linked to the characteristic sizes of the Q-balls, analogous to that of oscillon preheating in axion monodromy inflation \cite{Zhou:2013tsa}. The peak frequency of the GW power spectrum is also more carefully determined.

\section{Model and implementation}

There are typically many flat directions in supersymmetric extensions of the Standard Model, but the couplings between the flat direction fields themselves as well as the couplings of the flat direction fields to other light fields are usually suppressed \cite{Affleck:1984fy, Enqvist:1997si}.  Therefore, one may simply focus on one flat direction, which we will label as a complex scalar field $\Phi$ (with a canonical kinetic term). In gravity mediated SUSY breaking,  a typical effective potential for $\Phi$ is given by \cite{Enqvist:1997si,  Allahverdi:2012ju}
\bal
\label{poten}
V &= m_{3/2}^2 |\Phi|^2   + K m_{3/2}^2 |\Phi|^2 \log \frac{|\Phi|^2}{M_P^2}  
\nn
& ~~~ +  \left( Am_{3/2}\frac{\Phi^n}{M_P^{n-3}} +c.c.\right) +\lambda \frac{|\Phi|^{2n-2}}{M_P^{2n-6}}  ,
\eal
where $K, A, \lambda$ are dimensionless constants, $n$ is an integer, $M_P$ is the reduced Planck mass and $c.c.$ stands for complex conjugate. In this scenario, the flat direction (or sfermion) mass is linked to the mass scale of the gravitino. Considering the bounds from collider physics and naturalness, this mass may not be much far away from $\mc{O}({\rm TeV})$. The logarithmic term parameterizes the running of the flat direction mass and mainly comes from the contribution of one loop gaugino corrections, and typically the value of $K$ is roughly $-\alpha_s m^2_{1/2}/8\pi m^2_{\tilde{l}}$, where $m_{1/2}$ and $m_{\tilde{l}}$ are the gaugino and slepton masses respectively and $\alpha_s$ is the $SU(3)_c$ coupling constant  \cite{Allahverdi:2012ju,Dine:2003ax,Enqvist:2003gh}. The $A$ term and $\lambda$ term come from non-renormalizable interactions, with $A$ and $\lambda$ usually assumed to be order one. 
In the MSSM, $n$ is bigger than 3 and can be as big as 9 \cite{Gherghetta:1995dv}, or even bigger if further symmetries are assumed \cite{Dine:1995kz}. As mentioned in the introduction, the finite energy density in the early universe induces extra Hubble parameter dependent terms in the potential: $c_H H^2  |\Phi|^2$ and $\left( a_H H \Phi^n/M_P^{n-3} +c.c.\right)$, $c_H$ and $a_H$ being constants \cite{Dine:1995uk, Dine:1995kz}. The $c_H$ and $\lambda$ terms set the initial homogeneous value for $\Phi \simeq  M_P(m_{3/2}/M_P)^{1/(n-2)}$.

In the context of SUSY models, the $U(1)$ asymmetry generated are typically very small compared to the condensate itself \cite{Dine:1995uk, Dine:1995kz} but still typically bigger than the required baryon number density today $n_B/n_{\gamma}\sim 10^{-9}$. This can easily be solved by one of the following mechanisms \cite{Kusenko:2008zm, Kusenko:2009cv}: The $U(1)$ asymmetric terms are somehow suppressed; Or, there are substantial entropy productions at later stages; Or, the flat direction is that of $B-L= 0$ and the AD condensate decays above the scale where the active sphaleron process can destroy possible big $B+L$; Or, the flat direction is that of $B= 0$ and $L= 0$ \cite{Enqvist:2000cq, Enqvist:2002rj, Kusenko:2009cv}. GWs are only significantly produced when the AD condensate starts to fragment, after which time the $A$, $\lambda$, $c_H$ and $a_H$ terms become negligibly small. On the other hand, the GW production only depends on the energy momentum tensor of the condensate and is ``blind'' to the matter-anti-matter asymmetry in the AD condensate. So for our purposes we may safely neglect these terms,  which we have checked numerically. It is in this approximation that a global $U(1)$ symmetry emerges and Q-balls can be exactly defined.

Intuitively, Q-balls can form in this model because the AD condensate develops a negative pressure when the field oscillates around the minimum \cite{Kusenko:2009cv}. In other words, there is a process of parametric resonance, which  quasi-exponentially amplifies certain wavelengths of fluctuations. It follows that the AD condensate can efficiently fragment into lumps, which then evolve to spherically symmetric Q-balls. Technically, one may check that $V/|\Phi|^2$ (cf. Eq.~(\ref{poten}) with $K<0$, $A=0$ and $\lambda>0$) has a minimum at $|\Phi|\neq 0$, thus this potential is of the ``open-up'' type that supports Q-balls \cite{Coleman:1985ki}.

The ``open-up'' condition for Q-balls to form is very similar to that of oscillons \cite{Amin:2010jq}. Oscillons \cite{Bogolyubsky:1976yu} are also localized, spherically symmetric, time dependent field configurations, and can be produced from fragmentation of an inflaton condensate in preheating \cite{Amin:2011hj}. Unlike Q-balls, there is no Noether symmetry to guarantee their stability, but oscillons have a prolonged dissipation time. We will see shortly that Q-balls and oscillons play a very similar role in shaping the produced GW power spectrum.

For potential (\ref{poten}), $\ud^2 V/\ud |\Phi|^2$ has a soft, logarithmic divergence at $|\Phi|=0$.  Physically, small $|\Phi|$ is of course outside the validity of this effective potential, as the running mass correction becomes bigger than the leading term for small $|\Phi|$. Fortunately, this limitation of the effective potential does not significantly affect our computations. Note first that the equations of motion do not have any divergences. Also, the typical field value during its evolution is far away from the minimum. We find that numerically the effective mass only deviates slightly from $m_{3/2}$, by about $10\%$ in the simulations presented below. In the following, we will refer to the effective mass as $m_{3/2}$. 

Fragmentation of the AD condensate is a highly non-perturbative process and thus a numerical approach is needed to understand its full dynamics. Fortunately, a classical approximation is applicable to the fragmentation, as the occupation numbers for the relevant modes involved become very high once resonance kicks in. We assume the AD condensate becomes the dominant component of the universe before fragmentation \cite{Kusenko:2009cv}. We have run lattice simulations with metric perturbations included and found that the backreaction to the metric is as expected negligible, so it is sufficient to solve the Klein-Gordon equation in the Friedmann-Robertson-Walker (FRW) background:
\be
\label{kgV}
\bigg(\frac{\ud^2}{\ud t^2}+   3  H \frac{\ud}{\ud t}  -\frac{1}{a^2}\nd^2 \bigg)\Phi  =- \frac{\pd V}{\pd {\Phi}^*} ,
\ee
where $a$ is the scale factor, $H$ is the Hubble parameter, ${\Phi}^*$ is the complex conjugate of $\Phi$ and $\nd^2$ is the 3D flat space Laplacian. The evolution of the scale factor is determined by the Friedmann equation
\be
3M_P^2 H^2   = \left< \rho \right>  , ~~ \rho =  |\dot{\Phi}|^2  + |\nd \Phi|^2 +V   ,
\ee
where $\left<~\right>$ means averaging over the space. GWs are the tensor modes of the metric
\be
\ud s^2 =  -\ud t^2 + a^2(\delta_{ij} + h_{ij})\ud x^i \ud x^j    ,
\ee
where $h_{ij}$ is transverse and traceless, $\pd_i h_{ij} = h_{ii} =0$. To compute the stochastic GW background generated from fragmentation of the AD condensate, we solve the perturbed Einstein equations for the tensor modes (which are structurally similar to Eq.~(\ref{kgV})),
\be
\label{hijeom}
\left(\frac{\ud^2}{\ud t^2}+3H \frac{\ud}{\ud t} -\frac{1}{a^2}\nd^2 \right) h_{ij} = \frac{2}{M_P^2} T^{TT}_{ij}
\ee
where $T^{TT}_{ij}$ are the transverse and traceless part of the energy momentum tensor of the AD condensate. The energy density of the stochastic GW background is given by
\begin{align}
\label{rhogwdef}
\rho_{\rm gw} &= \frac{M_P^2}{4}  \left< \dot{h}_{ij} \dot{h}_{ij}  \right>
 \\
&= \frac{M_P^2}{4\mc{V}} \int\frac{\ud^3 k}{(2\pi)^3} \dot{h}_{ij}(t,\bfk) \dot{h}^{*}_{ij}(t,\bfk)  ,
\end{align}
where $\mc{V}$ is the space volume. Practically, one may solve an equation identical to Eq.~(\ref{hijeom}) except that $T^{TT}_{ij}$ is replaced by $T_{ij}$, the energy momentum tensor without the transverse and traceless projection, and, whenever an output is desired, do the projection at the level of Eq.~(\ref{rhogwdef}) in Fourier space, which is computationally much more efficient. One of the key observables we are interested in is the fractional GW energy density in the current epoch (we will refer to this quantity as the GW power spectrum in the following), which is defined by
\be
h^2 \Omega^0_{\rm gw}(f_0) = \frac{h^2}{\rho^0_c} \frac{\ud \rho^0_{\rm gw}}{\ud \ln f_0} = \frac{h^2}{\rho^0_c}  \frac{a^4}{a_0^4} \frac{\ud \rho_{\rm gw}}{\ud \ln k} 
\ee
where $h$ is the observational Hubble parameter today, $0$ denotes the current epoch of the universe, $f_0=k/2\pi a_0$ is the GW frequency today and $\rho^0_c=3M_P^2H_0^2$ is the critical energy density today.

Our lattice code is based on the public code HLATTICE \cite{Huang:2011gf}, where the equations of motion are re-cast in a different form in order to make use of more accurate, stable symplectic integrators. The use of a large lattice is particularly important in this study, because the separation of the lattice points become bigger with time, while the size of a Q-ball remains roughly the same. We need to make sure there are sufficient points to resolve a Q-ball at the end of the simulation. The results will be presented using a $256^3$ lattice, for which the GW power spectrum has sufficiently converged. The physical size of the lattice should be adjusted such that parametric resonance is captured in the simulations, leading Q-balls to form. This is achieved by choosing the lattice physical separation such that Q-balls are resolved within the lattice box. 

$m_{3/2}$ in the potential is chosen to be $1\,{\rm TeV}$ for the fiducial model, which applies to Figs.~\ref{rhoevo}, \ref{rhoevolution} and \ref{GWPSM}, but we will also consider $m_{3/2}=10\,{\rm TeV},100\,{\rm TeV}$ in Fig.~\ref{GW3m}, to see the dependence of the GW peak frequency on the SUSY scales. $K$ is typically within the range between $-0.01$ and $-0.1$ in MSSM \cite{Enqvist:1997si}, and we choose $K=-0.1$ as the fiducial value here. We find that the AD condensate fragments at a slower pace for a smaller $K$, but the total GW production is insensitive to $K$, consistent with the previous results with a smaller lattice ($64^3$) in Ref \cite{Kusenko:2009cv}. The initial value of the homogeneous mode $\Phi_0$ is chosen as $10^{16}\,$GeV, which corresponds to the case where the flat direction is lifted by $n=9$ terms in Eq.~(\ref{poten}).  The physical time scales of the fragmentation is insensitive to initial $\Phi_0$, but a smaller initial $\Phi_0$ does lead to a smaller scale factor $a$ at the onset of the fragmentation, as one would expect from the Friedmann equation. As the initial net charge is usually very small (A large net charge would be difficult to satisfy the current cosmological bounds on the matter-anti-matter asymmetry) and the GW production is blind to the U(1) charges, we will focus on a vanishing initial $\dot{\Phi}_0$ for simplicity. We have checked that adding some initial net charge has little impact on our main results. There are two kinds of inhomogeneities in the AD condensate that can be amplified by the parametric resonance process of the fragmentation: 1) Quantum fluctuations which exited the horizon during inflation and have re-entered the horizon afterwards; These inhomogeneities are roughly scale invariant, Gaussian and of order $|\delta\Phi/\Phi| \sim 10^{-5}$;  2) Quantum fluctuations of the AD condensate itself, which are also Gaussian but much smaller than the classicalized inhomogeneities from inflation. We will thus only include the first kind as the initial inhomogeneous seeds to our simulations, and consider an initial power spectrum for the complex field $\Phi$
\be
 \mc{P}_{\Phi/\Phi_0}(k) = \Delta_\mc{R}^2 \times \frac{2\pi^2}{k^3} =2.4\times 10^{-9} \times \frac{2\pi^2}{k^3}   .
\ee
(In comparison, the quantum fluctuation of the AD condensate itself is roughly $\mc{P}_{\Phi/\Phi_0}(k)\sim 1/\left(2\sqrt{m_{3/2}^2+k^2}|\Phi_0|^2\right)$, where $\Phi_0$ is the homogeneous mode mentioned above. This is much smaller at the time of fragmentation for all the relevant modes around $k\sim \mc{O}({\rm TeV})$.) However, we find that the total GW production is actually insensitive to the amplitude of the initial inhomogeneities, again consistent with the previous results with a smaller lattice ($64^3$) in Ref \cite{Kusenko:2009cv}.

\section{Numerical results}

\begin{figure}
\centering
\includegraphics[height=1.8in,width=3.3in]{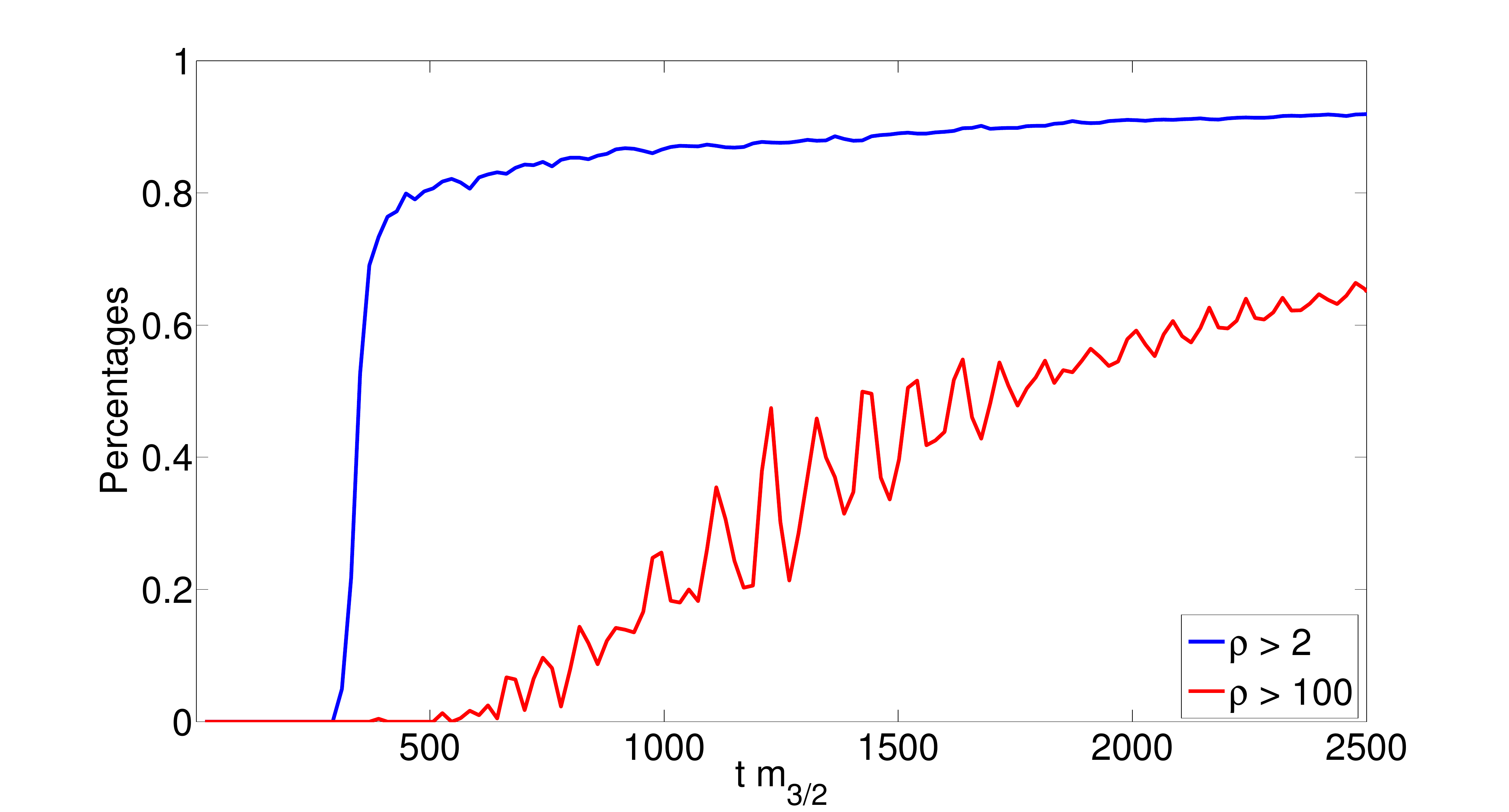}
\caption{Evolution of the percentage of the integrated energy in regions with an energy density $\rho>2$ (blue line) and $\rho>100$ (red line), for the fiducial model. The average energy density is normalized to 1. The lattice size is $256^3$, the initial box size equals $0.05H^{-1}$ and the lattice spacing and time step are $\ud x = 10 \ud t \simeq 0.078m^{-1}_{3/2}$.} 
\label{rhoevo}
\end{figure}

For our fiducial model, linear parametric resonance takes place before $t \sim 270m^{-1}_{3/2}$. After that, the dynamics becomes highly nonlinear and the AD condensate rapidly fragments into Q-balls. Q-balls have become the dominant component in the universe by $t\sim 520m^{-1}_{3/2}$ when the energy stored in regions with an energy density greater than twice the average reaches the plateau of about 80\%; See Fig.~\ref{rhoevo}. This plateau is only slowly lifted to about 90\% by the end of the simulation $t\sim 2500m^{-1}_{3/2}$ when Q-balls have properly formed. However, after domination, Q-balls continue to evolve towards spherical symmetry and become increasingly massive: 
After the fragmentation, the energy stored in regions with a density greater than 100 times the average increases almost linearly to about 65\% at the end of our simulation.

\begin{figure}[ht]
\centering
\subfigure[Linear perturbation]{%
\includegraphics[height=1.5in,width=1.4in]{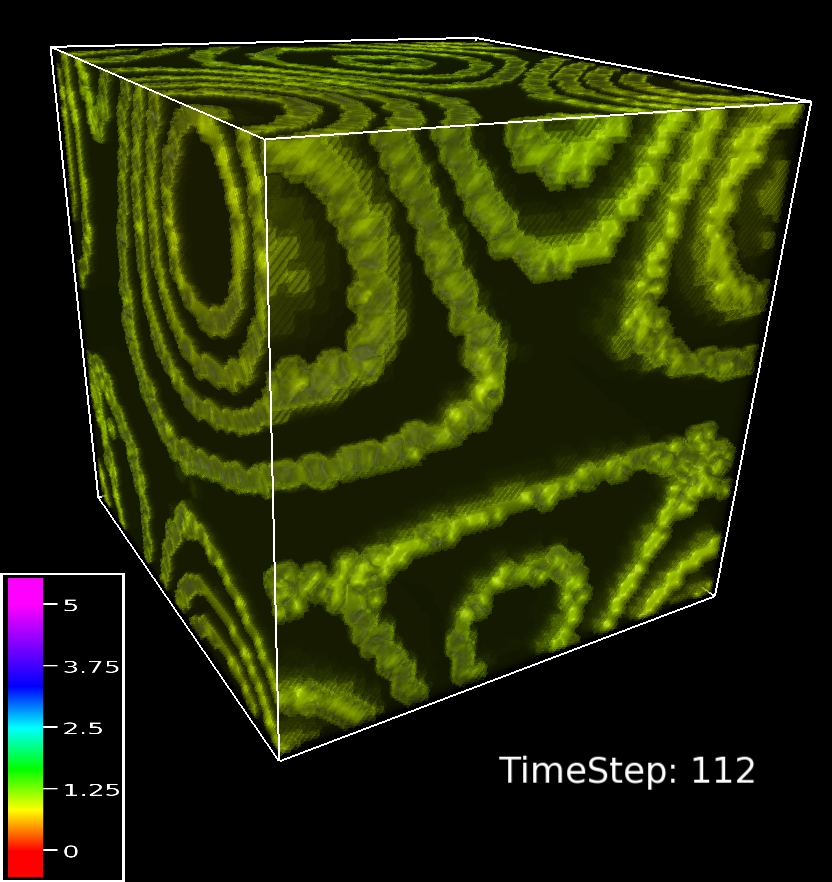}
\label{fig:subfigure1}}
\quad
\subfigure[Fragmentation]{%
\includegraphics[height=1.5in,width=1.4in]{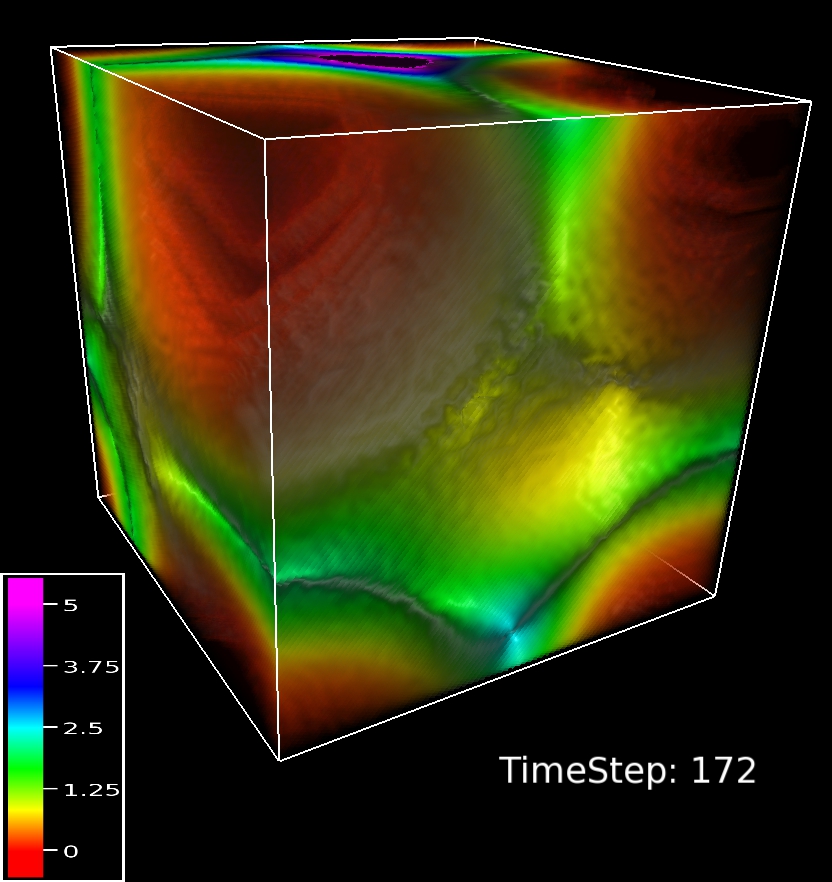}
\label{fig:subfigure2}}

\subfigure[Emerging Q-balls]{%
\includegraphics[height=1.5in,width=1.4in]{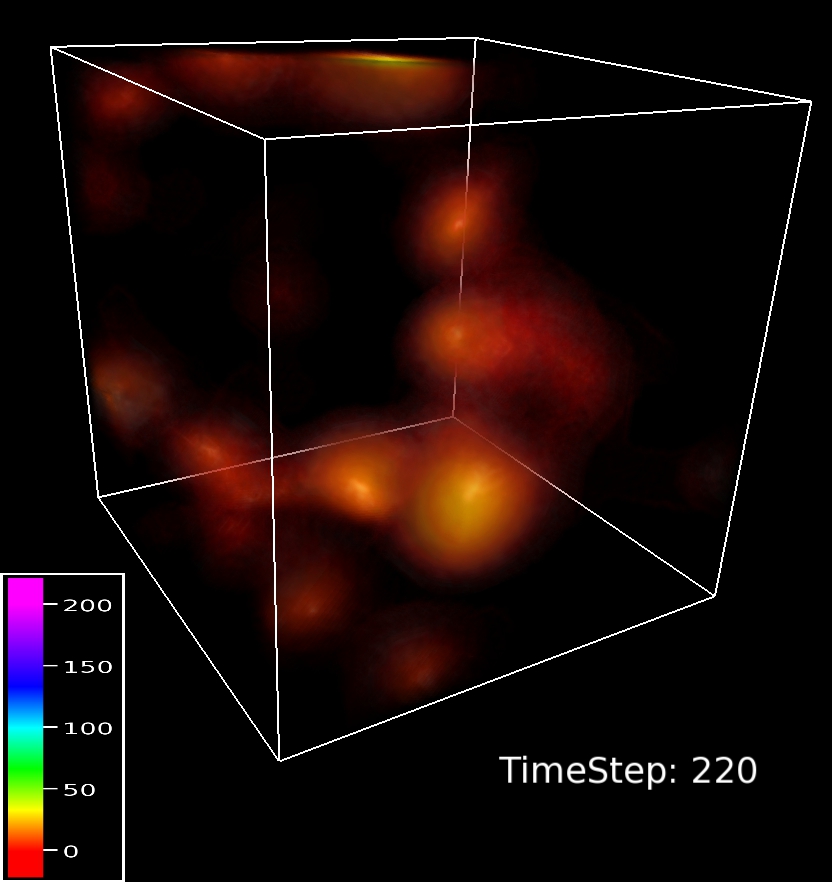}
\label{fig:subfigure3}}
\quad
\subfigure[Properly formed Q-balls]{%
\includegraphics[height=1.5in,width=1.4in]{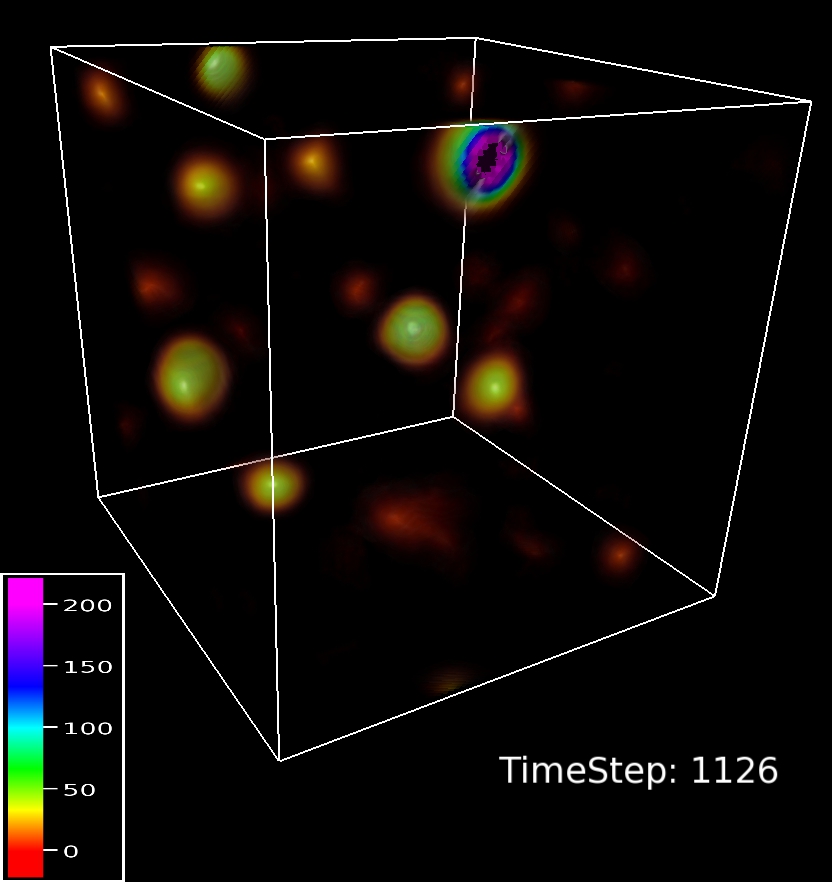}
\label{fig:subfigure4}}
\caption{Evolution of the energy density via volume rendering, from the same run as Fig.~\ref{rhoevo}. Note that the average energy density is normalized to 1, and the plotted max value of the energy density of (a) and (b) is 5, while for (c) and (d) it is 200. Lighting effects are used so that the minute linear perturbations in (a) can be seen. One ``TimeStep'' is $250 \ud t=1.95m^{-1}_{3/2}$. As can be seen from (d), a Q-ball is described by ${\cal O}(10^3)$ lattice points at the end of the simulation (box size: $256^3$).}
\label{rhoevolution}
\end{figure}

\begin{figure}
\centering
\includegraphics[height=2.1in,width=3.4in]{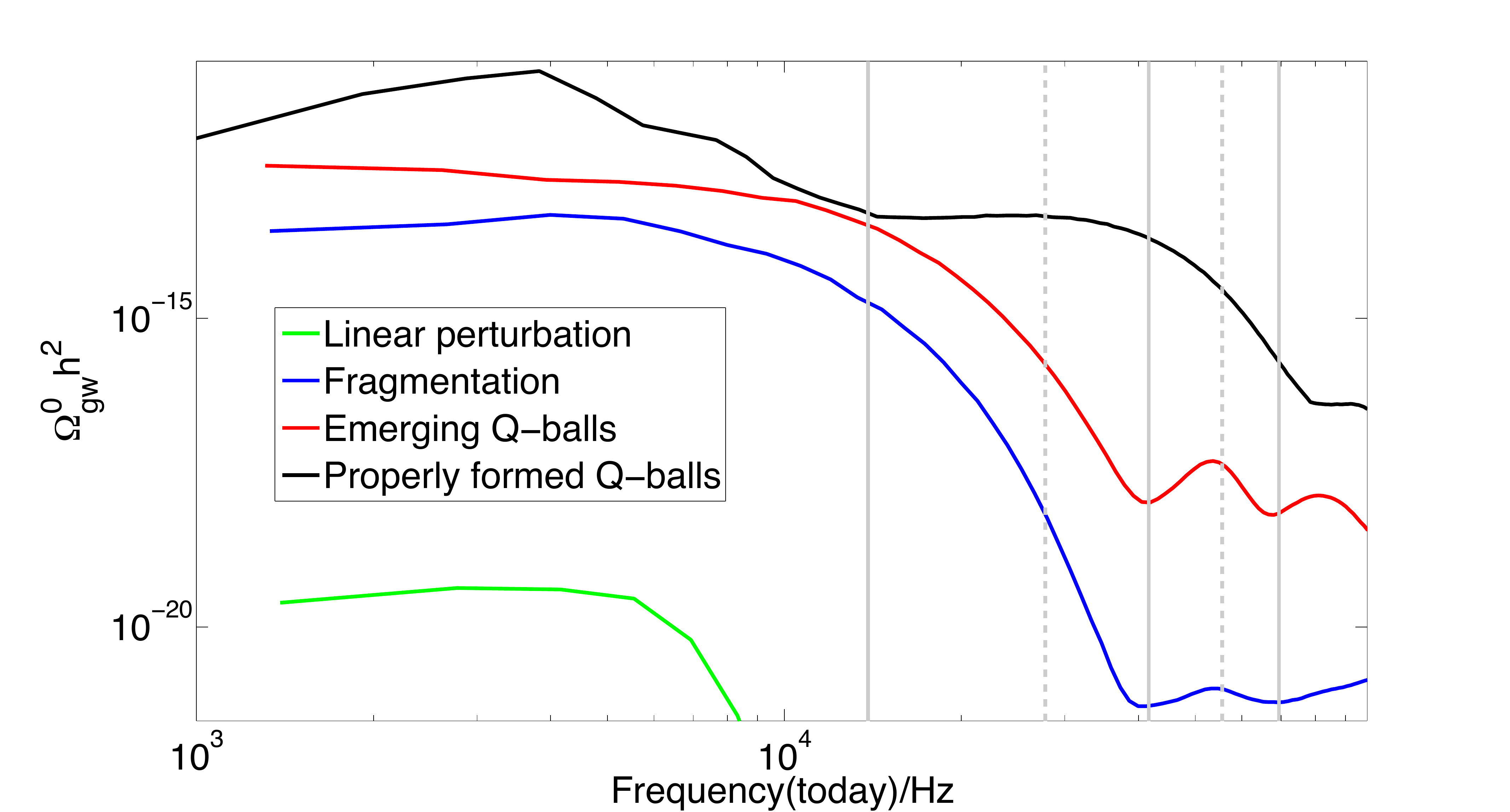}
\caption{Evolution of the GW power spectrum. The GW power spectrum grows upwards and the four spectra correspond to the four stages of Fig.~\ref{rhoevolution} respectively. The frequency modes in the bump of the blue line have grown due to parametric resonance.  The black line is the final stabilized GW power spectrum.  The vertical lines (solid and dashed) are the frequencies corresponding to $m_{3/2}$, $2m_{3/2}$, $3m_{3/2}$, $4m_{3/2}$, $5m_{3/2}$. 
The odd multiples of $m_{3/2}$ align with the troughs. This is analogous to the case of oscillon preheating \cite{Zhou:2013tsa}. The lattice size is again $256^3$, but a bigger box size $0.125H^{-1}$ is used to resolve the low frequencies. The lattice spacing and the time step are $\ud x = 10 \ud t \simeq 0.197m^{-1}_{3/2}$.} 
\label{GWPSM}
\end{figure}

We can divide the AD condensate fragmentation and its GW production into four stages (See Figs.~\ref{rhoevolution} and \ref{GWPSM}; the time scales are for the fiducial model.):
\begin{itemize}
\item
 Linear parametric resonance ($t\sim 0\to 270m^{-1}_{3/2}$): During this stage, the GW power spectrum grows steadily at the low momentum end of the spectrum. However, there is no multi-peak structure in the GW power spectrum yet.  The peak wavenumber of the linear growth of the perturbations has been estimated to be around $k_{\rm lin}^{2} \approx m_{\phi}^{2} |K| \left( 1- |K|/4\right)$~\cite{Enqvist:2002rj}, which agrees with our numerical results.
 \item
 AD condensate fragmentation ($t\sim 270m^{-1}_{3/2}\to 410m^{-1}_{3/2}$): This is when the fluctuations produced by linear parametric resonance has become sufficiently significant and the dynamics starts to become highly nonlinear. The AD condensate fragments efficiently during this short period. The low momentum modes of the fluctuations start to re-scatter to higher momentum modes. During this rapid process, the GW power spectrum grows very quickly, particularly at the higher momentum end of the spectrum. Towards the end of this stage, multiple peaks and troughs start to emerge in the GW power spectrum.
 \item
Q-balls emerging ($t\sim 410m^{-1}_{3/2} \to 520m^{-1}_{3/2}$): After the fragmentation, Q-balls emerge. The multiple peaks and troughs become apparent in the GW power spectrum. Most gravitational waves are produced during the first three stages, and we will refer to the duration of the third stages as the ``fragmentation time''.
\item
Q-balls become very massive ($t\sim 520m^{-1}_{3/2}\to$): After emerging, Q-balls slowly evolve towards spherical symmetry and steadily become more massive. The first trough slowly becomes increasingly apparent. The black curve in Fig.~\ref{GWPSM} is plotted around four folds of the fragmentation time when the GW power spectrum has become very stable.
\end{itemize}

In Fig.~\ref{GWPSM}, we see that the GW power spectrum at the frequency bands corresponding to the odd multiples of $m_{3/2}$ are suppressed, thus forming a multi-peak power spectrum. This is very much like that of oscillon preheating after axion monodromy inflation \cite{Zhou:2013tsa}, in which case the inflaton condensate fragments into oscillons. As the physical reason for this to happen in AD condensate fragmentation is almost identical to that in oscillon preheating, we shall only mention the essential points in the following (a detailed discussion of the completely analogous oscillon case can be found in \cite{Zhou:2013tsa}). Q-balls, like oscillons, formed out of a condensate's fragmentation, will evolve towards spherical symmetric lumps and spherically symmetric lumps do not emit GWs, so the GW frequency bands associated with the characteristic structures of Q-balls are suppressed in the spectrum. The characteristic structures of Q-balls are linked to the mass parameter, so Q-balls suppress the GW production at the frequency bands associated with the multiples of the mass. The reason why only frequencies linked to the odd multiples of the mass parameter are suppressed is due to the fact that the potential is invariant under $\Phi \to -\Phi$  \cite{Zhou:2013tsa}. However, the frequencies associated with the higher multiples are only suppressed during early times of Q-ball formation, since the higher frequency spherical structures decay at late times of the formation. So we end up with a GW power spectrum where the frequency band associated with the mass parameter itself (i.e., the first vertical line to the left, which corresponds to the overall size of the Q-ball) is suppressed. For the logarithmic Q-ball case at hand, the first trough linked to the Q-ball size appears relatively late and is relatively shallow, due to the softness of the logarithmic interaction.

\begin{figure}
\centering
\includegraphics[height=1.83in,width=3.4in]{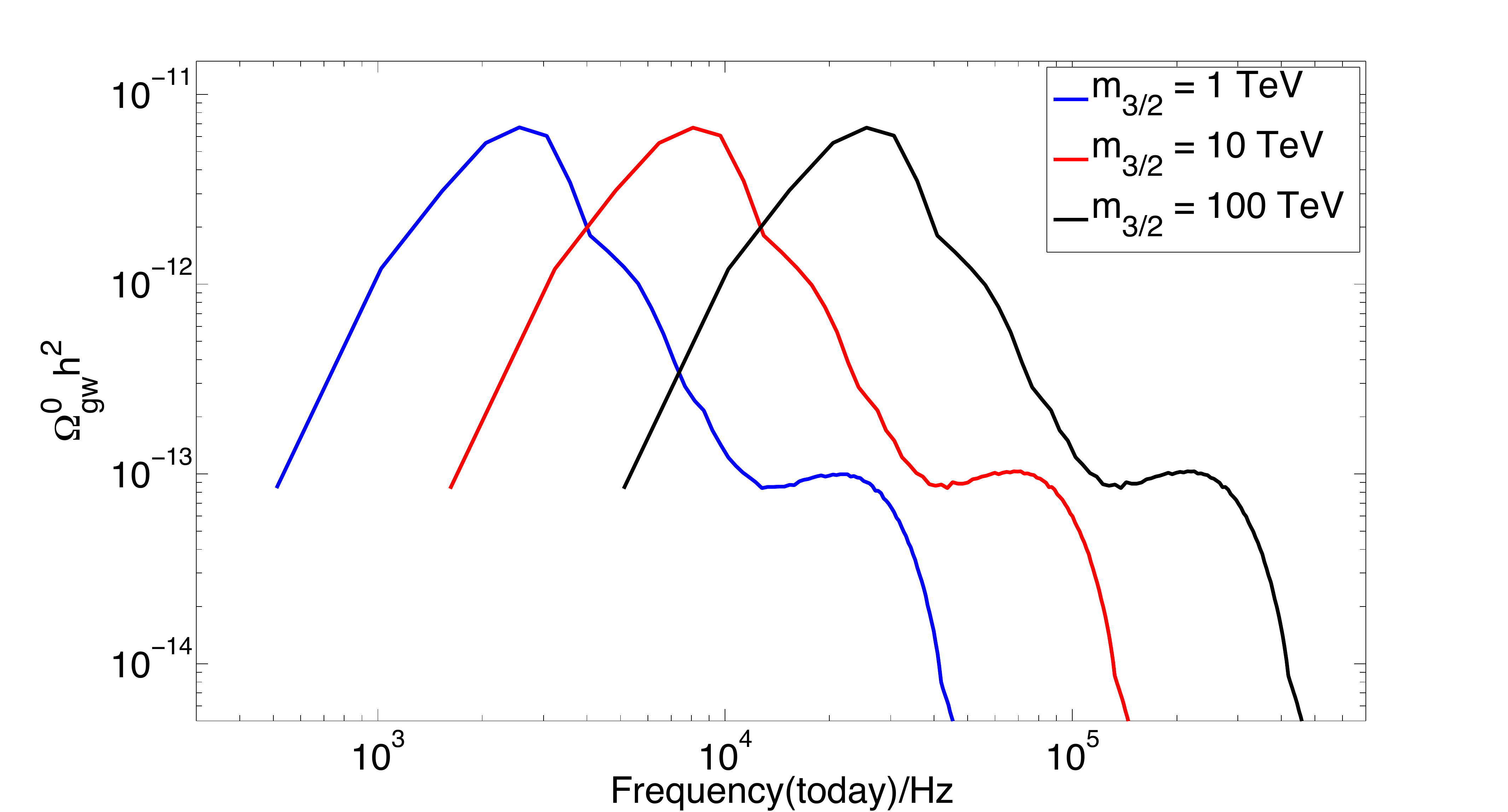}
\caption{Final GW power spectra for $m_{3/2}=1$\,TeV,  $10$\,TeV and  $100$\,TeV. The parameters and initial conditions are the same as the fiducial model, except for $m_{3/2}$. } 
\label{GW3m}
\end{figure}

The frequencies of the GW power spectrum (as observed today) have also been carefully determined in our simulations, which peaks around a few kHz for the case of $m_{3/2}= 1$ TeV (See Fig.~\ref{GWPSM}). Thus, the stochastic GW backgrounds for this scenario can not be observed by the current and upcoming interferometer based GW experiments (such as advanced LIGO)\;\footnote{To find out the sensitivity curves of the current and upcoming GW experiments, one can use the easy-to-use online utility: http://rhcole.com/apps/GWplotter/.}, contrary to the claims of \cite{Kusenko:2009cv}. However, new methods of observing ultra high frequency GWs are being developed, for example \cite{Arvanitaki:2012cn}. Since a bigger $m_{3/2}$ leads to higher GW frequencies, in the event of possible observations, one may use this stochastic GW background as a probe to the SUSY energy scale. See Fig.~\ref{GW3m} for the GW peak frequency's dependence on the SUSY scale $m_{3/2}$.
 
\section{Conclusion}

In summary, we have studied the dynamics of AD condensate fragmentation in the early universe and the associated stochastic GW production, based on the running mass potential (\ref{poten}) that typically arises from SUSY flat directions. Q-balls can be copiously produced in the AD condensate fragmentation, and a significant stochastic GW background can be produced when Q-balls are being formed out of the AD condensate. We have identified four distinct stages of this process. The generated GW power spectrum possesses a distinct multi-peak structure, which was unidentified in the previous studies. The trough in the GW multi-peak spectrum is linked to the SUSY breaking scale. This GW background, however, is not observable in the current or upcoming GW experiments (such as advanced LIGO), due to high peak frequencies and relatively small amplitudes. If this stochastic GW background can be observed in future high frequency GW detectors, one may use it as an independent probe to the SUSY energy scale. The SUSY energy scale this method can probe depends on how high the GW detector can probe a stochastic GW background's frequency. The reason for the trough to form in the GW spectrum is because spherically symmetric Q-balls are suppressing the GW production at the relevant frequencies, analogous to the scenario of oscillon preheating after axion monodromy inflation.

{\bf Acknowledgments}

We are grateful to Kari Enqvist, Hal Finkel, Zhiqi Huang, Andrea Romanino, Paul Saffin and Mitsuo Tsumagari for helpful discussions, and particularly thank Anupam Mazumdar for his collaboration in the early stage of this work. SYZ acknowledges support from DOE grant DE-SC0010600. The numerical work was performed on the HPC facilities in SISSA (hg1) and the CINECA Consortium (PLX). We also acknowledge the public software VAPOR for volume rendering in this work.

\end{document}